\documentclass[achemso,reprint,amsmath,amssymb,groupedaddress]{revtex4-1}

\usepackage{mathtools}
\DeclarePairedDelimiter{\ceil}{\lceil}{\rceil}
\DeclarePairedDelimiter{\floor}{\lfloor}{\rfloor}

\newcommand*{\C}{\mathbf{C}}
\newcommand*{\F}{\mathbf{F}}
\newcommand*{\I}{\mathbf{I}}
\newcommand*{\B}{\mathbf{B}}
\newcommand*{\A}{\mathbf{A}}
\newcommand*{\G}{\mathcal{G}}
\newcommand*{\vvec}{\mathbf{v}}
\newcommand*{\Rayleigh}{\vec{a}^t\cdot\F\cdot\vec{a}}
\newcommand*{\lm}{\lambda_{\text{min}}}
\newcommand*{\sgn}{\mathrm{sgn}}
\newcommand*{\cov}{\mathrm{cov}}
\newcommand*{\tr}{\mathrm{tr}}
\newcommand*{\eig}{\mathrm{eig}}
\newcommand*{\edges}{\mathrm{edges}}

\newtheorem{theorem}{Theorem}
\newtheorem{conjecture}{Conjecture}

\usepackage{graphicx}

\begin{document}

\title{Optimal measurement network of pairwise differences}

\author{Huafeng Xu}
\email[Correspondence to ]{huafeng@gmail.com}
\affiliation{Unaffiliated; Current address: Silicon Therapeutics, Boston, MA 02210, USA}

\begin{abstract}

When both the difference between two quantities and their individual
values can be measured or computationally predicted, multiple quantities
can be determined from the measurements or predictions of select
individual quantities and select pairwise differences.  These
measurements and predictions form a network connecting the quantities
through their differences.  Here, I analyze the optimization of such
networks, where the trace ($A$-optimal), the largest eigenvalue
($E$-optimal), or the determinant ($D$-optimal) of the covariance
matrix associated with the estimated quantities are minimized with
respect to the allocation of the measurement (or computational) cost
to different measurements (or predictions).  My statistical analysis
of the performance of such optimal measurement networks--based on
large sets of simulated data--suggests that they substantially
accelerate the determination of the quantities, and that they may be
useful in applications such as the computational prediction of binding
free energies of candidate drug molecules.

\end{abstract}

\maketitle

\section{Introduction}

It is quite common in a scientific study that multiple quantities are
to be determined from the measurements of their individual values and
their pairwise differences.  (I will refer to both experimental
measurements and computational predictions as measurements unless
otherwise distinguished.)  Compared to measuring only the individual
quantities, including measurements of pairwise differences may
substantially improve the statistical precision in the estimated
quantities.  This is especially true when the statistical
uncertainties in the measurements of the differences are much lower
than that in the measurements of the individual quantities, {\it
  e.g.}, when the differential value between two quantities lies
within the detection range of the experimental technique but the
individual values are outside the detection limit.

The statistical uncertainty associated with measuring a quantity or a
pairwise difference depends on the resources allocated to the
measurement, {\it i.e.} the number of repetitions in the case of
experimental measurements or the number of uncorrelated data points
sampled in the case of computational predictions.  Given fixed total
measurement resources ({\it i.e.} cost), optimal allocation of the
resources to different measurements may substantially improve the
overall statistical precision--characterized by the covariance
matrix--of the estimated quantities~\cite{Boyd2004,pukelsheim2006}.

An important example that may benefit from such optimal resource
allocations is the computational prediction of the binding affinities
of a set of molecules for a target receptor of pharmaceutical
interest.  Binding free energy calculations have demonstrated
sufficient accuracy~\cite{harder2016} and are increasingly adopted in
the pharmaceutical industry as an effective method to rank and select
candidate molecules in many drug discovery projects~\cite{wang2015}.
The binding free energy of an individual molecule can be computed by
the technique of absolute binding free energy
calculations~\cite{boresch2003,mobley2007,woo2005,aldeghi2015}, and
the difference between the binding free energies of any two molecules
can be computed by the technique of relative binding free energy
calculations~\cite{cournia2017,tembre1984,radmer1997,wang2015}.
Optimal estimators for {\em individual} absolute or relative binding free
energy calculations have been developed~\cite{bennett1976,shirts2008}.
At fixed computational cost, the statistical errors of free energy
calculations are bound by the thermodynamic length between the
thermodynamic end states~\cite{shenfeld2009}.  Typically, the end
states are more dissimilar in absolute binding free energy
calculations than in relative binding free energy calculations.  As a
result, the statistical errors in the former are often substantially
larger than those in the latter given the same amount of computational
cost.

The binding free energies of all the molecules in the set can be
estimated from the appropriate combinations of the predicted
individual values and the predicted pairwise differences (see below).
Optimal allocation of computational resources to different absolute
and relative binding free energy calculations may lead to improved
efficiency in predicting the binding free energies of a set of
molecules.  Past effort aimed at providing redundancy and consistency
checks in the calculations~\cite{liu2013}.  How to minimize the
overall statistical error in the estimated binding free energies,
however, remains an unaddressed question.

Here, I apply the mathematical results from optimal experimental
designs~\cite{Boyd2004,pukelsheim2006} to the special case of
optimizing the allocations of measurement resources to the
measurements of individual quantities and their pairwise differences,
so as to minimize the overall statistical error in the estimated
quantities.  Specifically, I consider three types of optimization: 1)
the $A$-optimal, which minimizes the trace of the covariance matrix and
hence the total variance, 2) the $D$-optimal, which minimizes the
determinant of the covariance matrix and hence the volume of the
confidence ellipsoid for a fixed confidence level, and 3) the
$E$-optimal, which minimizes the largest eigenvalue of the covariance
matrix and hence the diameter of the confidence ellipsoid.  The
$A$-optimal and the $D$-optimal are found by iterative numerical
minimization.  For the $E$-optimal, I present a new mathematical
theorem that enables the minimization to be solved by construction.  I
characterize the statistical performance of such optimizations in
terms of the reduction in the statistical error in the estimated
quantities at fixed total measurement cost. My results suggest that
optimal designs of the measurements can substantially reduce the
statistical error in the estimated quantities, allowing the same
statistical precision (characterized by the total statistical error)
to be achieved at--on average--less than half the measurement cost
when compared to naive allocations.  The Python code for generating
the optimal designs is made available as free open source software
(https://github.com/forcefield/DiffNet).

\section{Theory}

Suppose that we want to measure a set of quantities $\{
x_{i=1,2,\dots,m} \}$.  We can either measure each individual quantity
$x_i$ with the estimator $\hat{x}_i = x_i + \sigma_i$, or the
difference between any pair of quantities $x_i$ and $x_j$ with the
estimator $\hat{x}_{ij} = x_i - x_j + \sigma_{ij}$, where $\sigma_i$
and $\sigma_{ij}$ are the respective statistical errors in the
measurements.  For each measurement $e\in \{i|i=1,2,...,m\} \cup \{
(i,j)|i,j=1,2,\dots,m, i\neq j\}$, the statistical variance of the
estimate decreases with $n_e$--the resource allocated to the
measurement--as
\begin{equation}
\sigma_e^2 = s_e^2/n_e
\label{eqn:variance-at-n}
\end{equation}
where $s_e$ is the statistical fluctuation in the corresponding
experimental measurement or computer sampling.  

The quantities and the measurements form a network--which I will refer
to as the {\em difference network}--represented by a graph $\G$ of
$m+1$ vertices and $m(m+1)/2$ edges, where the vertices
$i=1,2,\dots,m$ stand for the $m$ quantities $\{ x_i \}$, the edge
between vertices $i\neq j>0$ stands for the difference measurement
$\hat{x}_{ij}$, and the edge between vertices $0$ and $i>0$ stands for
the individual measurement $\hat{x}_i$.  Two weighted graphs can be
derived from $\G$: 1) $\G_s$, in which the weight assigned to each edge
is the corresponding fluctuation $s_e$, and 2) $\G_n$, in which the
weight is the corresponding number of samples $n_e$.  I will denote
$s_{i,0} = s_{0,i} = s_i$ and $n_{i,0} = n_{0,i} = n_i$.

For a given set of $\{ n_e \}$ and the corresponding $\{ \sigma_e \}$
per Eq.~\ref{eqn:variance-at-n}, the maximum likelihood estimator for
$\{ x_i \}$ is~\cite{Boyd2004,pukelsheim2006,wang2013} (assuming that
the statistical errors in the measurements follow the normal
distribution; see Appendix~\ref{sec:ML-proof} for a derivation)
\begin{equation}
\F\cdot \vec{x} = \vec{z}
\label{eqn:ML-estimator}
\end{equation}
where
\begin{equation}
z_i = \sigma_i^{-2}\hat{x}_i + \sum_{j\neq i} \sigma_{ij}^{-2} \hat{x}_{ij},
\label{eqn:ML-estimator-RHS}
\end{equation}
and $\F$ is the Fisher information matrix:
\begin{equation}
F_{ij} = \left\{
\begin{array}{rl}
\sigma_i^{-2}+\sum_{k\neq i}\sigma_{ik}^{-2} & \text{ if  } i=j \\
-\sigma_{ij}^{-2} & \text{ if  } i\neq j 
\end{array}
\right..
\label{eqn:Fisher-information}
\end{equation}

The covariance in the estimates of $\{ x_i \}$ is given by the inverse of
the Fisher information matrix:
\begin{equation}
\C = \F^{-1}
\label{eqn:covariance}
\end{equation}

In the classical theory of optimal design of
experiments~\cite{pukelsheim2006}, three objectives of minimizations
of statistical errors with respect to $\{ n_e \}$ are commonly sought:
\begin{itemize}

\item $A$-optimal: minimize $\tr(\C)$,
\item $D$-optimal: minimize $\ln\det(\C)$,
\item $E$-optimal: minimize $||\C||_2$,

\end{itemize}
where $\tr(\C)$ denotes the trace of the matrix $\C$, $\det(\C)$
denotes the determinant of $\C$, and $||\C||_2$ denotes the spectral
norm--or, equivalently, the largest eigenvalue--of $\C$, subject to
the constraints of non-negativity
\begin{equation}
n_e \geq 0,
\label{eqn:non-negativity}
\end{equation}
and of total fixed measurement cost
\begin{equation}
\sum_e n_e = \sum_i n_i + \sum_{i<j} n_{ij} = N.
\label{eqn:constant-resource}
\end{equation}

Although $\{ n_e \}$ are integers, the above minimizations are more
conveniently performed if $\{ n_e \}$ are allowed to be real numbers.
They can then be rounded to integers
(see Appendix.~\ref{sec:rounding-to-int}).  This is a reasonable
approximation when $N \gg m(m+1)/2$.  This condition
is usually satisfied in sampling-based computational predictions such
as binding free energy calculations, because in such cases $N$ is the
total number of independent sampling points, which can be in the range
of tens of millions.  Automated experiments permitting large numbers
of repetitions may also satisfy this condition.  

All three objectives are convex functions of $\{ n_e \}$ (see
Appendix~\ref{sec:convexity-proof}), and the minimization can be
performed by standard algorithms in convex
optimization~\cite{Boyd2004}.  For the $A$-optimal design, the
minimization can be solved as a semidefinite programming problem (SDP)
(see Appendix~\ref{sec:SDP-A-optimal}).

For the $E$-optimal design, the minimization can be similarly cast as
an SDP.  But for the difference network, I present the following
theorem (see Appendix~\ref{sec:E-tree-proof} for proof), which allows
the minimization problem to be solved by construction, with a fixed
time complexity of $O(m^2)$ and space complexity of $O(m)$, and which,
to my knowledge, has not appeared in any previous publication.

\begin{theorem}
Let $E_i$ be the shortest path from vertex 0 to vertex $i$ in the
graph $\G_s$ (i.e., the path with the smallest $\sum_{e\in E_i} s_e$),
and $E = \cup_i E_i$ be the tree rooted at vertex 0 from the resulting
union.  (A tree is a connected acyclic undirected graph.)  Denoting
$a_0=0$ and
\begin{equation}
a_i = \sum_{e\in E_i} s_e \text{ for } i>0,
\end{equation}
$||\C(\{ n_e \})||_2$ is minimized by the following set of $\{ n_e \}$:
\begin{equation}
n_{i\mu_i} = N s_{i\mu_i} \sum_{j\in T_i} a_j \left(\sum_i a_i^2\right)^{-1}
\label{eqn:E-tree-optimal}
\end{equation}
where $T_i \subset E$ is the subtree rooted at vertex $i$, and $\mu_i$
is the vertex immediately preceding $i$ in the path $E_i$ from 0 to
$i$.  
\label{thm:E-tree}
\end{theorem}

The shortest paths $\{ E_i \}$ can be constructed by the
single-source-multiple-destination Dijkstra algorithm (see, {\it
  e.g.}, Ref.~\onlinecite{Mehlhorn2008}), which guarantees that
$E=\cup_i E_i$ is a tree.  The $E$-optimal by construction is
substantially faster than by minimization using SDP: for random $\{
s_{e} \}$ drawn uniformly in the interval $(1, 5)$, the speedup of the
former over the latter (implemented in CVXOPT~\cite{CVXOPT}) is $\sim
400\times$ for $m=10$ and $\sim 3700\times$ for $m=50$.  In
Appendix~\ref{sec:constant-rel-e}, I suggest a special class of
difference networks in which the $A$- and the $D$-optimals may also be
solved by construction.

Often, there is a cost $\tau_e$ associated with generating each
sampling point in the estimator $\hat{x}_e$, and the total cost is
$\sum_e n_e \tau_e = N$. We can, however, introduce $n_e^\dagger=n_e
\tau_e$, and $s_e^\dagger = s_e \tau_e^{1/2}$, and solve the same
problem as above for $n_e^\dagger$, with the parameters
$s_e^\dagger$.

\section{Results}

\begin{figure*}
\center
\includegraphics[width=\textwidth]{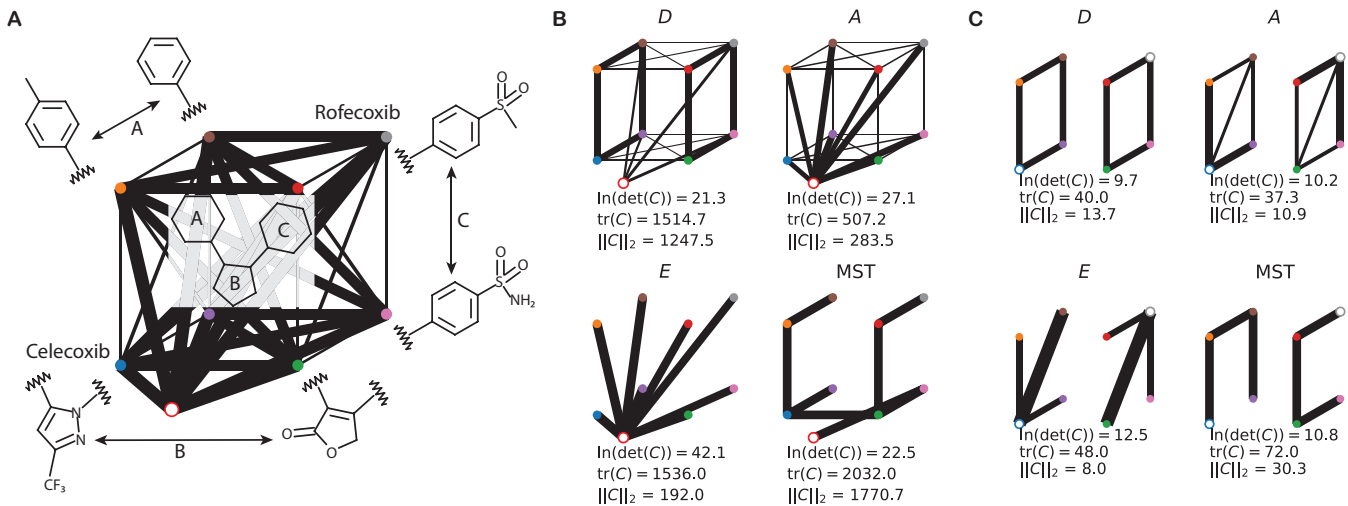}
\caption{ Optimal allocation of computational sampling to the binding
  free energy calculations for COX-2 inhibitors. \textbf{A} The 8
  inhibitors are generated by different combinations of 3 rings, with
  2 different options for each ring.  The weighted graph $\G_s$ is
  shown.  Each filled vertex represents an inhibitor.  The edge
  between any two such vertices $i$ and $j$ represents the relative
  binding free energy calculation between inhibitors $i$ and $j$, and
  the thickness of the edge is proportional to the corresponding
  $s_{ij}$.  The edge between the unfilled vertex at the bottom and a
  vertex $i$ represents the absolute binding free energy calculation
  of inhibitor $i$, and its thickness is proportional to
  $s_i$. \textbf{B} The $D$-, $A$-, and $E$-optimals of the
  corresponding difference network, where the thickness of each edge
  is proportional $n_e$ according to the optimal
  allocations. \textbf{C} The optimals when the binding free energies
  of Celecoxib and Rofecoxib are known and used as references, and
  only relative binding free energy calculations are used to computed
  the binding free energies of the other 6 inhibitors.  For
  comparison, the minimum spanning tree (MST) of $\G_s$ is also shown,
  where $n_e$ is constant for each edge in the tree.  }
\label{fig:COX-2}
\end{figure*}

I first illustrate the optimal difference networks using the example
of the calculation of the binding free energies of $m=8$ inhibitors
for the COX-2 protein~\cite{yamakawa2014}.  The optimals depend on the
statistical fluctuations $\{ s_e \} = \{ s_i | i=1,2,\dots, m \} \cup
\{ s_{ij} | i,j=1,2,\dots,m, i<j\}$. For illustrative purposes, I
set $\{ s_e \}$ such that $s_i$ in the absolute binding
free energy calculations is $s_{i} = \zeta \sqrt{h_i}$, where $h_i$ is
the number of heavy atoms in molecule $i$ and $\zeta$ is a constant,
and that $s_{ij}$ in the relative binding free energy calculations is
$s_{ij} = \zeta \sqrt{\max( h_{ij}, h_{ji} )}$, where $h_{ij}$ is the
number of heavy atoms in molecule $i$ that do not transform into atoms
in molecule $j$ in the relative binding free energy calculation of
molecules $i$ and $j$.  The corresponding $\G_s$ is shown in
Fig.~\ref{fig:COX-2}\textbf{A}.  In real binding free energy
calculations, the $\{ s_e \}$ depends on both the thermodynamic length
between the end states~\cite{shenfeld2009} and the ratio of the
relaxation time of relevant motions to the length of the simulation;
they have to be determined (approximately) during iterative rounds of
binding free energy calculations.  The optimal allocations
corresponding to my hypothetical $\{ s_e \}$ are shown in
Fig.~\ref{fig:COX-2}\textbf{B}.

In a typical drug discovery projects, there will be a few molecules
whose binding free energies have already been experimentally
determined, and, using these molecules as references, relative binding
free energy calculations can be used to predict the binding free
energies for other similar molecules.  The $D$-, $A$-, and
$E$-optimals for such relative binding free energy calculations are
shown in Fig.~\ref{fig:COX-2}\textbf{C} for the COX-2 inhibitors,
using Celecoxib and Rofecoxib as the reference molecules.  Because the
$\{ s_e \}$ values above reflects the assummption that relative
binding free energy calculations have substantially lower statistical
errors than absolute binding free energy calculations, this network of
relative binding free energy calculations--taking advantage of the
known binding free energies of the reference molecules--yield much
lower overall statistical errors than the previous network in
Fig.~\ref{fig:COX-2}\textbf{B}.

Next, I characterize the statistical performance of the optimal
difference networks in comparison to naive choices of $\{ n_e \}$.
The $A$-, $D$-, $E$-optimals and various other naive allocations are
applied to randomly generated sets of $\{ s_e \}$, and the resulting
covariance matrices are compared in terms of their traces,
determinants, and spectral norms.    

\begin{figure}
\includegraphics[width=3in]{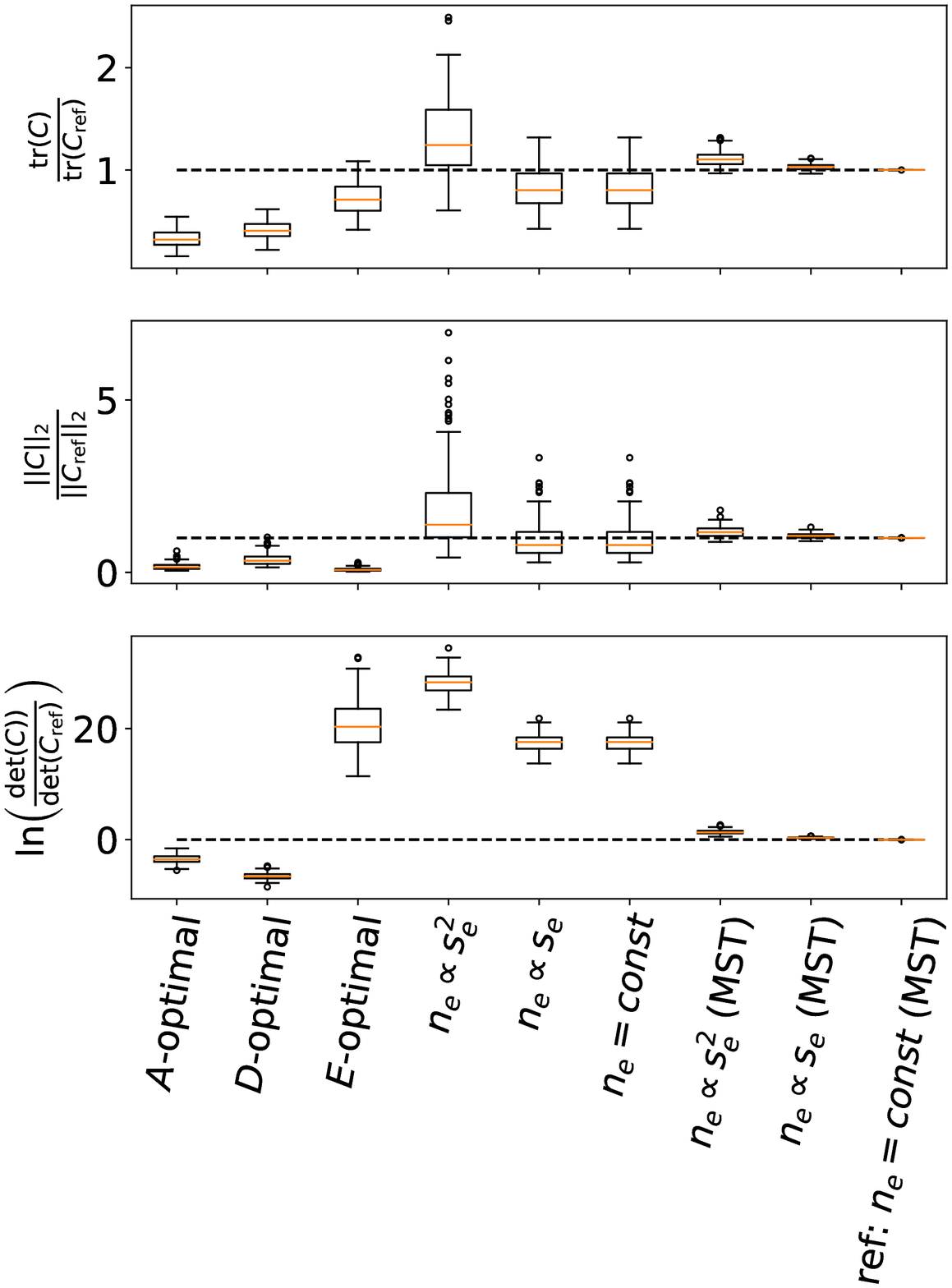}
\caption{Statistical performance of different allocations.  Here
  $m=30$, $\{ s_{ij} \}$ and $\{ s_i \}$ are uniformly disitrbuted
  between 1 and 5.  As the reference, I chose $n_e = const$ for $e \in
  \text{MST}(\G_s)$, where $\text{MST}(\G_s)$ is the minimum spanning
  tree (MST) of the weighted graph $\G_s$.  The covariances
  of the various allocations are computed for $T=200$ randomly
  generated sets of $\{ s_e \}$, and the ratios of $\tr(\C)$,
  $||\C||_2$, and $\det(\C)$ to the corresponding reference values are
  shown here in boxplots, with the median ratios and the quartiles
  indicated. }
\label{fig:gain-random}
\end{figure}

For the sets of random $\{ s_e \}$ drawn uniformly from the interval
of $(1, 5)$, the $A$-optimal outperforms all tested schemes of naive
allocations by all three criteria ($\tr(\C)$, $||\C||_2$, and
$\det(\C)$; it also yields a $||\C||_2$ close to that of the
$E$-optimal and a $\ln(\det(\C))$ close to that of the $D$-optimal
(Fig.~\ref{fig:gain-random}).  Compared to the $D$-optimal, which has
the second lowest average $\tr(\C)$, the $A$-optimal reduces $\tr(\C)$
by a factor of $0.791\pm 0.005$ ({\it i.e.} the ratio of $\tr(\C)$ for
the $A$-optimal to that for the $D$-optimal is $0.791$); compared to
the naive allocation $n_{e}\propto s_{e}$, which has the lowest
average $\tr(\C)$ among the tested naive allocations, the $A$-optimal
reduces $\tr(\C)$ by a factor of $0.402\pm 0.001$.  These observations
suggest that the $A$-optimal--which has the simple interpretation of
minimizing the total variance of the measured quantities--may be a
good default choice in designing difference networks.

The $A$-optimal enables substantial improvements in statistical
precision for difference networks with other distributions of $\{ s_e
\}$ as well.  For example, in the case where $s_i = s_{ij} =
\text{constant}$ and $m=30$, the $A$-optimal reduces $\tr(\C)$ by a factor
of 0.664 compared to the naive allocation $n_e \propto s_e$; in the
case where the relative error is constant (such that $s_{ij} = |s_i -
s_j|$; see Appendix~\ref{sec:constant-rel-e}) and $m=30$ values of
$s_i$ are drawn randomly from the interval $(0, 1)$, the corresponding
reduction is by a factor of $0.470\pm 0.001$.

\begin{figure}
\includegraphics[width=3.5in]{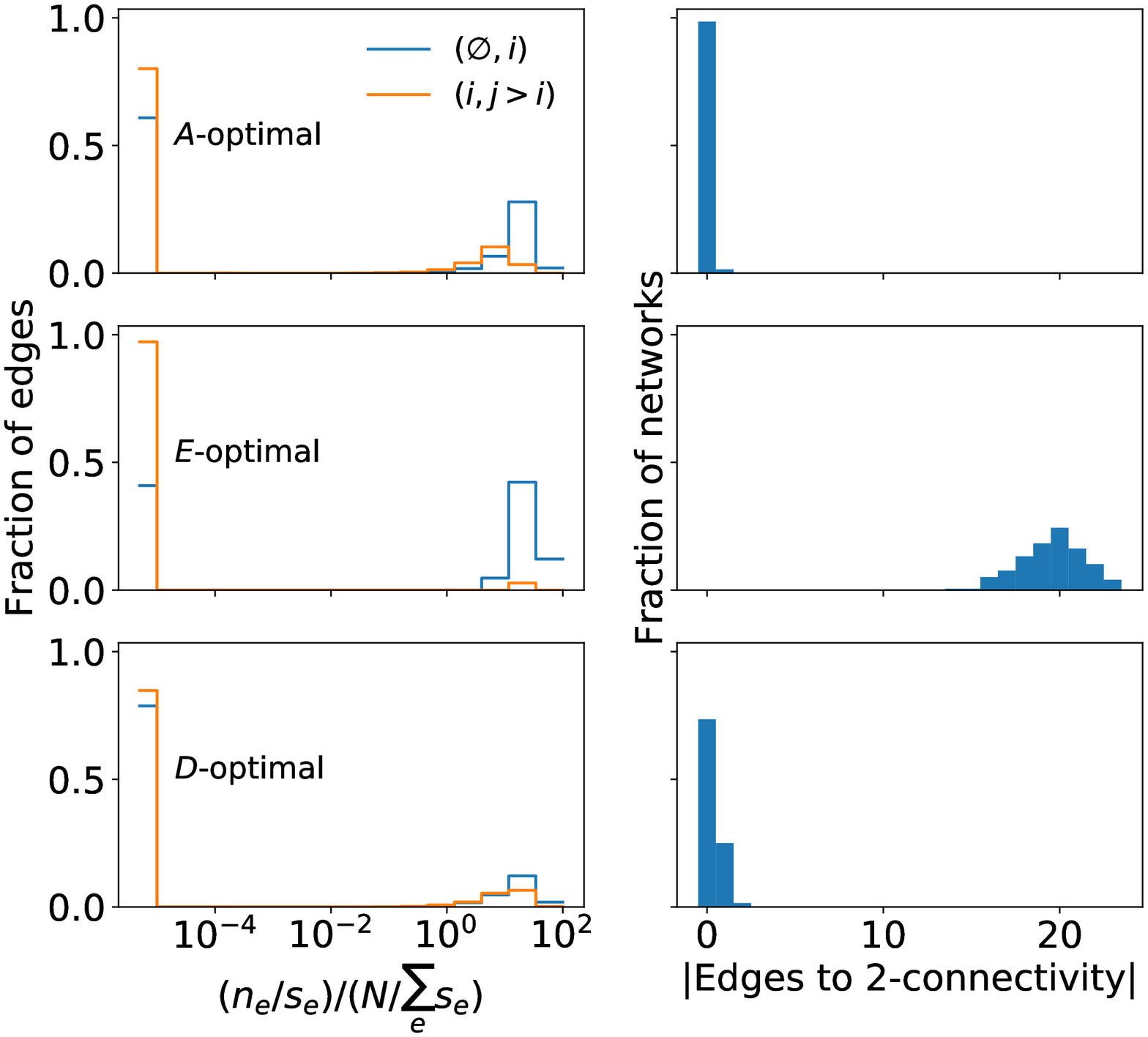}
\caption{Statistical distributions of $\{ n_e \}$ in the optimal
  allocations, for the same random sets of $\{ s_e \}$ as in
  Fig.~\ref{fig:gain-random}.  The fractions of edges in the weighted
  graph $\G_n$ with different $n_e/s_e$ ratios are shown on the left.
  The right panels show the distributions of the minimum number of
  edges that need to be added to the graph $\G_n$--after eliminating
  the edges with negligible allocations (determined by
  $n_e/s_e/(N/\sum_e s_e) < 10^{-2}$)--to make it 2-connected ({\it
    i.e.} there are at least 2 distinct paths connecting any two
  vertices in $\G_n$).  The majority of the $A$-optimal networks are
  2-connected. }
\label{fig:nij-random}
\end{figure}

It is desirable that the measurement networks are 2-connected--a
$k$-edge-connected subgraph is one that does not become disconnected
unless $k$ edges are removed--such that for any two quantities, there
are two paths through which their differences can be computed.  This
allows a self-consistency check: the differences computed both ways
should be approximately equal.  Such checks can reveal potential
measurement errors and outliers.  The majority (98.5\%) of the
$A$-optimal networks are 2-connected, and all of them become
2-connected with the addition of at most 1 edge
(Fig.~\ref{fig:nij-random}).  This property of the $A$-optimal
allocation also makes it a good choice in designing difference
networks.

The $A$- and $D$-optimal networks may be densely connected (the
$E$-optimal is a tree hence sparsely connected).  For example, in the
$A$-optimal allocations for the above randomly generated sets of $\{
s_e \}$, on average individual measurements are applied to 11.8 of the
$m=30$ individual quantities, and difference measurements are applied
to 86.7 of the $m(m-1)/2=435$ pairs (Fig.~\ref{fig:nij-random}).  The
$A$- and $D$-optimal difference networks for the COX-2 binding free
energy calculations (Fig.~\ref{fig:COX-2}) are also clearly dense.  As
suggested in Appendix~\ref{sec:constant-rel-e}, however, the $A$- and
$D$-optimal networks are trees when $s_{ij} = |s_i - s_j|$.

Practical considerations--such as the minimum number of samples
required in an individual binding free energy calculation to derive
meaningful free energy estimates, and how many measurements can be
executed in parallel--often make it desirable to limit the number of
measurements to include in the difference network, {\it i.e.} to limit
the number of edges in $\G_n$.  Appendix~\ref{sec:sparse-diffnet}
outlines a heuristic approach to designing near-optimal difference
networks given the required edge-connectedness $k$ and the number of
measurements $M$ (the number of edges in $\G_n$ with $n_e > 0$).
The resulting network, $\G_{k,M}$, can be further pruned to eliminate
any edge with $n_e/N < \epsilon/M$ ({\it e.g.}, to avoid impractically
short binding free energy calculations; $\epsilon$ is a
parameter that specifies the cutoff).  When applied to the $A$-optimal
of the example difference network with randomly generated fluctuations
(described in Fig.~\ref{fig:gain-random}), the $\tr(\C)$ of the
near-optimal allocation generated by this approach, with $k=2$,
$M=3m$, and $\epsilon=0.1$, is on average only $1.10\pm 0.03$ times
that of the true optimal.

Design of the optimal difference network requires as input the
statistical fluctuations $\{ s_e \}$ associated with each measurement,
which is usually unknown {\it a priori}.  Thus the application of the
optimal difference networks needs to proceed in an iterative manner.
Starting with an initial guess of $\{ s_e \}$ ({\it e.g.} the
statistical fluctuation in relative or absolute binding free energy
calculations may be predicted by a machine-learning model trained on a
collection of past free energy calculations), the quantities can be
measured with the corresponding optimal network at a small total
measurement cost; For any measurement $e$ performed, update $s_e$ by
the actual statistical variance of the measurement, and re-optimize
the network accordingly; Repeat this process at increasingly larger
total cost until sufficiently precise estimates of the quantities are
reached.

\begin{figure}
\includegraphics[width=3in]{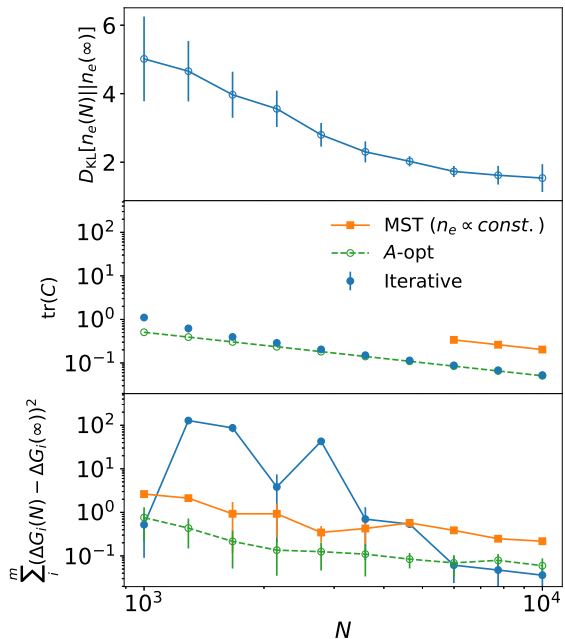}
\caption{Iterative $A$-optimization of the difference network. Here,
  the difference network corresponds to that of binding free energy
  calculations for the eight COX-2 inhibitors, and the true
  fluctuations $\{ s_e^\infty \}$ are taken to be the same as given in
  Fig.~\ref{fig:COX-2}\textbf{A}.  Top: The Kullback-Leibler
  divergence, $D_{\text{KL}}$, between the allocation in each
  iteration $n_e(N)$ and the optimal allocation $n_e(\infty)$
  according to the true $\{ s_e^\infty \}$ at different total number
  of samples $N$ after each iteration.  Middle: $\tr(C)$ versus $N$
  for the iterative optimization, compared to that for the true
  $A$-optimal and the minimal spanning tree allocation according to
  $\{ s_e^\infty \}$.  Bottom: The square deviation of the estimated
  free energies $\Delta G_i(N)$ after each iteration from the true
  free energies $\Delta G_i(\infty)$.  No actual free energy
  calculations were performed.  Instead, $\Delta\Delta G_{ij}$ and
  their fluctuations $s_{ij}$ are calculated as the average and
  standard deviations of $n_{ij}$ random numbers drawn from a normal
  distribution with the mean $\Delta G_i(\infty) - \Delta G_j(\infty)$
  and the standard deviation $s_{ij}^\infty$.  $\Delta G_i$ are
  estimated from $\Delta\Delta G_{ij}$ and $s_{ij}^2/n_{ij}$ using
  Eq.~\ref{eqn:ML-estimator}.  Iterative optimization quickly
  approaches the true optimal; at $N=10^4$, $\tr(\C)$ of the
  iteratively optimized difference network is practically
  indistinguishable from that of the true $A$-optimal, and it is $0.249
  \pm 0.004$ times that of the minimal spanning tree allocation.  }
\label{fig:diffnet-iter}
\end{figure}

Fig.~\ref{fig:diffnet-iter} illustrates this iterative process in the
construction of the $A$-optimal for a difference network. In the
beginning, $\{ s_e \}$ were initialized to be random numbers drawn
uniformly from the interval of $(0, 1)$.  In each iteration, $s_e$ was
updated to be the actual estimate of the fluctuation for any performed
measurement $e$ ({\it i.e.} $n_e > 0$), or to be a random number drawn
uniformly between the minimum and the maximum of all the estimated
fluctuations ({\it i.e.}  in the interval $(\min_{e'|n_{e'}>0}(\{
s_{e'} \}), \max_{e'|n_{e'}>0}(\{ s_{e'} \})$) if measurement $e$ has
not yet been performed ({\it i.e.} $n_e = 0$).  In each iteration, a
prescribed total of $\Delta N = \sum_e \Delta n_e$ additional samples
are allocated so as to optimize $\tr(\C)$ with respect to $\Delta n_e$
(see Appendix~\ref{sec:iteration}), and $\Delta n_e$ are rounded to
integers according to Appendix~\ref{sec:rounding-to-int}.  In a few
iterations the allocation $\{ n_e(N) \}$ becomes close to the optimal
allocation, $\{ n_e(\infty)\}$, according to the true fluctuations $\{
s_e^\infty \}$.  The Kullback-Leiber divergence from $\{ n_e(N) \}$ to
$\{ n_e(\infty) \}$
\begin{equation}
D_{\text{KL}}[n_e(N)||n_e(\infty)] = 
\sum_e \frac{n_e(N)}{N} \ln \frac{n_e(N)}{n_e(\infty)}
\label{eqn:KL-divergence}
\end{equation}
decreases in each iteration.  Correspondingly, the total statistical
error in the estimated quantities approaches the true minimum at
sufficiently large $N$.

\section{Discussions}

This work explored the application of optimal experimental design to
the determination of multiple quantities by the measurements of select
individual quantities and select pairwise differences, such as in the
computational prediction of binding free energies for candidate drug
molecules.  A related work was recently published by Yang {\it et
  al.}~\cite{Yang2019}, in which the authors proposed using discrete
$A$- or $D$-optimals to select the pairs of relative binding free
energy calculations for predicting the binding free energies of a set
of molecules.  Whereas Yang {\it et al.} addressed the question of
between {\em which} pairs of molecules relative binding free energy
calculations should be performed, given a fixed {\em number of
  calculations} ({\it i.e.}  $n_e=0,1$, under the constraint $\sum_e
n_e = N$, $N$ being the total number of calculations), I addressed
here the question of {\em how much} sampling should be allocated to
the relative binding free energy calculations between each possible
pair of molecules, given a fixed total {\em amount of sampling} ({\it
  i.e.} $0\le n_e \le N$, under the constraint $\sum_e n_e = N$, $N$
being the total number of samples in all the calculations).  The
expanded domain of $\{ n_e \}$ in the latter allows additional control
in the design of the calculations, which further reduces the resulting
statistical errors.

The optimal difference network may help accelerate a variety of
scientific inquiries.  Improving the statistical precision of binding
free energy predictions, for instance, allows better selection of
candidate drug molecules in drug discovery projects, as suggested by
Yang {\it et al.}~\cite{Yang2019}.  The optimal difference network may
find immediate use in the parametrization of computational models,
where the same set of quantities are computed repeatedly in search for
the values of the model parameters that best fit benchmark data.  For
example, binding free energy calculations for sets of inhibitors
binding to their receptor targets~\cite{harder2016} and solvation free
energy calculations for a large number of solutes~\cite{mobley2014}
are routinely used to test the accuracy of molecular force fields.
The increased statistical efficiency conferred by the optimal
difference networks will allow faster assessment of the quantitative
accuracy of the models and thus shorten their development cycles.

\appendix

\section{Derivation of the estimator and the covariance in the difference network}
\label{sec:ML-proof}

The set of quantities $\{ x_i \}$ can be estimated from the
measurements $\{ \hat{x}_e \}$ by the maximum likelihood (ML)
estimator.  Let $\sigma_e^2 = s_e^2/n_e$ be the variance for 
the measurement $e$, the likelihood that the measurements produce
the set of values $\{ \hat{x}_e \}$, given the true values of the
quantities $\{ x_i \}$, is
%
\begin{eqnarray}
L &=& \prod_i (\sqrt{2\pi}\sigma_i)^{-1}\exp\left(-\frac{(x_i - \hat{x}_i)^2}{2\sigma_i^2}\right) 
\nonumber \\
&\cdot& \prod_{i<j}(\sqrt{2\pi}\sigma_{ij})^{-1}\exp\left(-\frac{(x_i - x_j - \hat{x}_{ij})^2}{2\sigma_{ij}^2}\right)
\label{eqn:likelihood}
\end{eqnarray}
The above assumes that the samples in the measurements are independent.

Maximizing $L$ with respect to $\{ x_i \}$ yields
Eq.~\ref{eqn:ML-estimator}-\ref{eqn:Fisher-information}.  The
convariance matrix for the ML estimator is given by the inverse of the
Fisher information matrix, hence Eq.~\ref{eqn:covariance}.

\section{Rounding $n_e$ to integers}
\label{sec:rounding-to-int}

In practice $n_e$ need to be rounded to integers that sum to
$N$. Let $E_+ = \{ e|n_e > 0 \}$ be the set of
measurements requiring samples, and $|E_+|$ be the
size (or cardinality) of $E_+$. One can sort $E_+$ by
increasing values of $n_e$ (such that $n_{e_j}$ is the
$j$th lowest value), and round up the lowest $k$ values and round down
the highest $|E_+| - k$ values of $n_e$ to the nearest
integers. $k$ is chosen such that
\begin{equation} 
\sum_{j=1}^k \ceil{n_{e_j}} + \sum_{j=k+1}^{|E_+|} \floor{n_{e_j}} = N,
\end{equation}
which yields
\begin{equation}
k = N - \sum_{e\in E_+} \ceil{n_e} - |E_+|.
\end{equation}

\section{Proof of convexity of $\tr(\C)$ and $||\C||_2$}
\label{sec:convexity-proof}

The objective functions $\tr(\C)$, $\det(\C)$, and $||\C||_2$ have all
been previously shown to be convex functions of $\{ n_e
\}$~\cite{Boyd2004}.  Here I include simple proofs for the convexity
of $\tr(\C)$ and $||\C||_2$.

The Fisher information matrix $\F = \C^{-1}$ is a linear function of
$\{ n_e \}$.  Both $\F$ and $\C$ are symmetric and positive definite.
Consider a perturbation $q\A$ to $\F$, where $\A$ is an arbitrary
symmetric matrix.  The perturbed covariance matrix satisfies
\begin{equation}
\C(q)\cdot(\F + q\A) = (\F + q\A)\cdot\C(q) = \I
\label{eqn:CqFq}
\end{equation}
where $\I$ is the identity matrix.  Differentiating Eq.~\ref{eqn:CqFq}
with respect to $q$ twice, we have
\begin{equation}
\frac{d^2}{dq^2} \C(q) = 2(\A\cdot\C(q))^t\cdot\C(q)\cdot(\A\cdot\C(q))
\end{equation}
which is positive definite.

Thus 
\begin{equation}
\frac{d^2}{dq^2} \tr(\C(q)) = \tr\left(\frac{d^2}{dq^2}\C(q)\right) > 0
\end{equation}
proving the convexity of $\tr(\C)$ in $\{ n_e \}$.

To prove that $||\C||_2$ is convex in $\{ n_e \}$, note that
\begin{equation}
||\C||_2 = \lambda_1(\F)^{-1}
\label{eqn:spectral-norm}
\end{equation}
where $\lambda_1(\F) \equiv \min\{\eig(\F)\}$ is the smallest eigenvalue
of $\F$. 

According to the second-order perturbation theory, for the perturbed matrix
$\F + q\A$, the corresponding $\lambda_1(q)$ is
\begin{equation}
\lambda_1(q) = \lambda_1(\F) + q \vvec_1^t\cdot A \cdot \vvec_1
 + q^2\sum_{k=2}^m \frac{\left(\vvec_k^t\cdot A \cdot \vvec_k\right)^2}{\lambda_1 - \lambda_k} + O(q^3)
\end{equation}
where $\lambda_k$ is the $k$'th eigenvalue of $\F$ ($k=1,2,\dots,m$),
and $\vvec_k$ is the corresponding normalized eigenvector. The second
derivative of $\lambda_1$ is
\begin{equation}
\frac{d^2}{dq^2}\lambda_1 = \sum_{k=2}^m \frac{\left(\vvec_k^t\cdot A \cdot \vvec_k\right)^2}{\lambda_1 - \lambda_k} \leq 0
\end{equation}
because $\lambda_1 \leq \lambda_k$, 

Thus
\begin{equation}
\frac{d^2}{dq^2}||\C||_2 = \frac{2\left(d\lambda_1/dq\right)^2 - \lambda_1 d^2\lambda_1/dq^2}{\lambda_1^3} \geq 0
\end{equation}
proving the convexity of $||\C||_2$ in $\{ n_e \}$.

\section{Semidefinite programming for $A$-optimal design}
\label{sec:SDP-A-optimal}
For the $A$-optimal design, the
minimization can be cast by duality as a semidefinite programming
problem (SDP):
\begin{eqnarray}
& \text{minimize } & \sum_{i=1}^{m} u_i \nonumber \\
& \text{subject to } & 
\left(
\begin{array}{cc}
\F(\{n_e\}) & \mathbf{e}_i \\
\mathbf{e}_i^t & u_i \\
\end{array}
\right) \succeq 0 \text{ for $i=1,2,\dots,m$}, \nonumber \\
&& n_e \geq 0 \text{ for all }e, \text{ and } \sum_e n_e = N, 
\label{eqn:dual-A-optimal}
\end{eqnarray}
where $\mathbf{M}\succeq 0$ signifies that the symmetric matrix
$\mathbf{M}$ is positive semidefinite, and $\mathbf{e}_i$ is the $i$th
unit vector ({\it i.e.} $e_{ij} = \delta_{ij}$).

\section{Proof that Eq.~\ref{eqn:E-tree-optimal} is the $E$-optimal}
\label{sec:E-tree-proof}

The largest eigenvalue of $\C$ is the inverse of the smallest
eigenvalue of $\F$ (Eq.~\ref{eqn:spectral-norm}), so the problem of
minimizing $||\C||_2$ is equivalent to maximizing the smallest
eigenvalue of $\F$:
\begin{equation}
\max_{\{n_e\}} {\min\{ \eig(\F) \}}
\end{equation}

The smallest eigenvalue $\lm$ of $\F$ satisfies
\begin{equation}
\lm \leq \Rayleigh
\end{equation}
for all normal vectors $\vec{a}, |a|^2=1$; the equality holds if and only if
$\vec{a}$ is the eigenvector of $\F$ corresponding to $\lm$. Our problem is
thus to find
\begin{equation}
\max_{\{ n_e \}} \min_{|a|^2=1} \Rayleigh
\label{eqn:maxmin}
\end{equation}

I have
\begin{eqnarray}
\Rayleigh &=& \sum_i\sigma_i^{-2} a_i^2 
  + \sum_{j\neq i}\sigma_{ij}^{-2} a_i^2 
  - \sum_{j\neq i}\sigma_{ij}^{-2} a_i a_j \nonumber \\
&=& \sum_i\sigma_i^{-2} a_i^2 + \sum_{i<j}\sigma_{ij}^{-2}(a_i-a_j)^2 .
\end{eqnarray}

Plugging in $\sigma_e^{2} = s_e^2/n_e$, I have
\begin{equation}
\Rayleigh = \sum_i\frac{n_i a_i^2}{s_i^2} 
          + \sum_{i<j}\frac{n_{ij} (a_i - a_j)^2}{s_{ij}^2}.
\label{eqn:Rayleigh}
\end{equation}

Given any vector $\vec{a}$, eq.~\ref{eqn:Rayleigh} is maximized with
respect to $\{ n_e\}$ when the only non-zero $n_e$'s are the ones
corresponding to the largest value of $a_i^2/s_i^2$ or
$(a_i-a_j)^2/s_{ij}^2$. There may be degenerate set of $E = \{ e \}$ with the
largest values:
\begin{equation}
\left.\frac{a_i}{s_i}\right|_{i\in E} = \left.\frac{|a_i - a_j|}{s_{ij}}\right|_{ij \in E} = R_m \geq \left.\frac{a_i}{s_i}\right|_{i\not\in E}, \left.\frac{|a_i - a_j|}{s_{ij}}\right|_{ij \not\in E}
\label{eqn:max_ratio}
\end{equation}
and
\begin{equation}
\max_{ \{ n_e \} } { \Rayleigh } = \sum_e n_e R_m = N R_m
\end{equation}
The maximimum can be achieved by different values of $\{ n_e \}$, so long as $n_e \neq 0$ only if $e\in E$. 

If I can determine a set $E$ and a set of $\{ n_e^\ast
\}$ such that $n_e^\ast \neq 0 \iff e\in E$, and the corresponding
Fisher information matrix $\F(\{n_e^\ast\})$ has an eigenvector
$\vec{a}$ of the lowest eigenvalue satisfying eq.~\ref{eqn:max_ratio},
I have
\begin{equation}
\vec{a}^t\cdot\F(\{n_e^\ast\})\cdot\vec{a} \geq \vec{a}^t\cdot\F(\{ n_e \})\cdot\vec{a} \geq \min_{|a'|^2=1} \vec{a}'^t\cdot\F(\{ n_e \})\cdot\vec{a}' 
\end{equation}
Such a set of $\{ n_e^\ast \}$ would thus be a solution of
eq.~\ref{eqn:maxmin}. There may be degenerate solutions of
eq.~\ref{eqn:maxmin} and to our problem.

Next I show how to construct such a set of $\{ n_e \}$ by
constructing its corresponding eigenvector $\vec{a}$ that satisfies
eq.~\ref{eqn:max_ratio}.

Consider a complete graph $\G_s$ consisting of $N+1$ vertices, indexed
as $0,1,2,\dots,m$. The edge between the vertex $i$ and the vertex $0$
is assigned weight $s_i$, and the edge between the vertex $i$ and the
vertex $j$ is assigned weight $s_{ij}$. For each vertex $i\neq 0$,
find the shortest path $E_i$ from $0$ (i.e., with the smallest sum
$a_i = \sum_{e\in E_i} s_e$). We also denote $a_0 = 0$,
$s_{i,0}=s_{0,i}=s_i$, and $n_{i,0}=n_{0,i}=n_i$.  I will show that
the vector $\vec{a} = \{ a_{i=1,2,\dots,m} \}$ is the sought
eigenvector, and that $E$ comprises the edges of a tree--rooted at
vertex $0$--that is the union of the shortest path to each vertex
$i>0$.

Consider the union of $\{ E_i \}$: $E' = \cup_i E_i$.  If there is any
loop in $E'$, an arbitrary edge in the loop can be removed and the
resulting graph will still contain the shortest path from $0$ to every
vertex $i$.  Thus the tree $E$ can be constructed by removing all the
loops in $\cup_i E_i$.  In fact, if $\{E_i\}$ are found by the
Dijkstra's algorithm, $\cup_i E_i$ will not contain any loops.  This
is taken to be the case below.
 
First, I prove that $\vec{a}$ and the corresponding $E$ satisfy
eq.~\ref{eqn:max_ratio}. Clearly 
\begin{equation}
\left.\frac{|a_i - a_j|}{s_{ij}}\right|_{ij \in E} = 1
\end{equation}
If there exists a pair $i,j$ such that $(a_i - a_j)/s_{ij} > 1$, it
implies that the shortest path to $i$ is $E_j$ followed by edge $ji$,
with the sum $a_j + s_{ij} < a_i$, which is a contradiction.

I can now construct the set $\{ n_e \}$ so that $\vec{a}$ is the
eigenvector of $\F(\{ n_e \})$, i.e., 
\begin{equation}
\F(\{ n_e \}) \vec{a} = \lambda \vec{a}.
\label{eqn:Fn-eigen}
\end{equation}

Note that $n_e \neq 0$ only if $e \in E$. Denote the set of vertices whose edge
to vertex $i$ are in $E$ as $J_i$ (which may include the vertex 0),
the elements of $\F$ can be written as
\begin{equation}
F_{ii} = \sum_{j\in J_i}\frac{n_{ij}}{s_{ij}^2}
\end{equation}
and
\begin{equation}
F_{ij} = \left\{
\begin{array}{rl}
-\frac{n_{ij}}{s_{ij}^2} & \text{if } j\in J_i \\
0 & \text{otherwise}  
\end{array}
\right.
\end{equation}
and Eq.~\ref{eqn:Fn-eigen} becomes
\begin{eqnarray}
\lambda a_i 
&=& F_{ii} a_i + \sum_{j\neq i,\  j>0} F_{ij} a_j \nonumber \\
&=& \sum_{j\in J_i} \frac{n_{ij}}{s_{ij}^2} a_i - \sum_{j\in J_i\setminus\{0\}} \frac{n_{ij}}{s_{ij}^2} a_j \nonumber \\
&& (\because a_0 = 0) \nonumber \\
&=& \sum_{j\in J_i} \frac{n_{ij}}{s_{ij}^2} (a_i - a_j) \nonumber \\
&& (\because j\in J_i \iff (i,j)\in E \Rightarrow |a_i - a_j| = s_{ij}) \nonumber \\
&=& \sum_{j\in J_i} \frac{n_{ij}}{s_{ij}}\sgn(a_i-a_j)
\end{eqnarray}

Let $\mu_i$ be the vertex immediately preceding $i$ in the path $E_i$
connecting $0$ to $i$. $j=\mu_i$ is the only vertex in $J_i$ for which
$a_i > a_j$. Otherwise, let $j'$ be another vertex such that $a_i >
a_{j'}$, which implies that the edge $(ij')$ cannot be part of
$E_{j'}$, and thus both $E_i$ and $E_{j'}\cup{(ij')}$ are paths
connecting $0$ to $i$, in contradiction to the fact that $E$ is
without any loop. Another corollary is that the edge $(i\mu_i)$ must
be in the path $E_j$ for every $j$ in the subtree $T_i \subset E$
rooted at vertex $i$.

I can write
\begin{equation}
\frac{n_{i\mu_i}}{s_{i\mu_i}} = \lambda a_i + \sum_{j\in J_i\setminus\{ \mu_i \}}\frac{n_{ij}}{s_{ij}}
\label{eqn:n_recursion}
\end{equation}

Eq.~\ref{eqn:n_recursion} can be solved by starting from the leaves of
the tree $E$ and working backwards toward vertex $0$. The solution is
\begin{equation}
\frac{n_{i\mu_i}}{s_{i\mu_i}} = \lambda \sum_{j\in T_i } a_j
\label{eqn:n_expression}
\end{equation}
where the sum runs over the set of vertices in the subtree $T_i$
rooted at $i$, including $i$ itself.

The eigenvalue $\lambda$ can be determined by applying the constraint
$\sum_e n_e = \sum_i n_{i\mu_i} = N$.
\begin{widetext}
\begin{eqnarray}
\sum_i n_{i\mu_i} &=& \lambda \sum_i s_{i\mu_i} \sum_{j\in T_i} a_j
\nonumber \\
&=& \lambda \sum_i (a_i - a_{\mu_i})\left(a_i + \sum_{j\in T_i\setminus\{i\}} a_j\right)
\nonumber \\
&=& \lambda\left( \sum_i a_i^2 + \sum_i a_i \sum_{j\in T_i\setminus\{i\}} a_j - \sum_i a_{\mu_i} \left(a_i + \sum_{j\in T_i\setminus\{i\}} a_j\right) \right)
\end{eqnarray}
\end{widetext}

In both $\sum_i a_i\sum_{j\in T_i\setminus\{i\}} a_j$ and $\sum_i
a_{\mu_i}(a_i + \sum_{j\in T_i\setminus\{i\}} a_j)$, the product $a_i
a_{j\neq i}$ appears once and only once if and only if $j$ is in the subtree
of $T_i$ or $i$ is in the subtree of $T_j$. This implies that the two
sums are equal, and I have
\begin{equation}
N = \sum_i n_{i\mu_i} = \lambda \sum_i a_i^2
\end{equation}
or
\begin{equation}
\lambda = N \left(\sum_i a_i^2\right)^{-1}
\label{eqn:lambda}
\end{equation}
 
To prove that this $\lambda$ in eq.~\ref{eqn:lambda} is the smallest
eigenvalue of $\F$, I show that it is the largest eigenvalue of the
corresponding covariance matrix $\C=\F^{-1}$.

$x_i$ is estimated--because $E$ is a tree rooted at $0$--by
\begin{equation}
x_i = \sum_{e \in E_i} \hat{x}_e
\end{equation}
thus the covariance between $x_i$ and $x_j$ is
\begin{eqnarray}
C_{ij} &=& \sum_{e \in E_i, e'\in E_j} \cov(\hat{x}_e, \hat{x}_{e'})
\nonumber \\
&=& \sum_{e \in E_i\cap E_j} \sigma_e^2 
= \sum_{e \in E_i\cap E_j} \frac{s_e^2}{n_e}
\label{eqn:E-covariance}
\end{eqnarray}

Without loss of generality, let's assume that within $E$ vertices
$1,2,\dots,b$ have edges to $0$. $\C$ can be rearranged into block
diagonal form where each block $\B_{k=1,2,\dots,b}$ corresponds to the
covariances between pairs of vertices both within the subtree
$T_k$. Any eigenvalue $\C$ must be the eigenvalue of $\B_k$. Since
$C_{ij}>0$ for any two vertices sharing edges in $E_i$ and $E_j$, each
$\B_k$ is a positive matrix, and by Perron-Frobenius theorem it has
only one eigenvector with all positive components, and the
corresponding eigenvalue has the largest magnitude. Since $\vec{a}$ is
an eigenvector of $\C$ with all positive components, its corresponding
eigenvalue $(\sum_i a_i^2)/N$ must be the eigenvalue of the largest
magnitude for each $\B_k$, thus be the eigenvalue of the largest
magnitude for $\C$. Q.E.D.

\section{The special case of constant relative error}
\label{sec:constant-rel-e}

In the special case where the relative error for any measurement is
the same, {\it i.e.}, $s_e/\hat{x}_e$ is a constant, $D$- and
$A$-optimals may also be solved by construction.  Without loss of
generality,  the quantities can be ordered by $x_1 \leq x_2 \leq \dots
\leq x_m$ (which is often easily done by qualitative comparison
without quantitative measurement), and it is apparent that $s_{ij} =
s_j - s_i$ for $j>i$.  I conjecture that in such cases of constant
relative error, the $D$- and $A$-optimals can be solved by the
following construction:

\begin{conjecture}
If $s_i \leq s_j$ and $s_{ij} = s_j - s_i$ for $i<j$, $\det(C)$ is minimized by 
\begin{eqnarray}
n_1 &=& N/m \nonumber \\
n_{i>1} &=& 0 \nonumber \\
n_{i\,i+1} &=& N/m \nonumber \\
n_{ij} &=& 0 \text{ if } |j-i| > 1
\label{eqn:D-optimal-const-rel-e}
\end{eqnarray}
\label{cnj:D-optimal-const-rel-e}
\end{conjecture}

\begin{conjecture}
If $s_i \leq s_j$ and $s_{ij} = s_j - s_i$ for $i<j$, $\tr(C)$ is minimized by 
\begin{eqnarray}
n_1 &=& \lambda \sqrt{m}\cdot s_1 \nonumber \\
n_{i>1} &=& 0 \nonumber \\
n_{i\,i+1} &=& \lambda \sqrt{m-i} \cdot (s_{i+1} - s_i) \nonumber \\
n_{ij} &=& 0, \text{ if } |j-i| > 1
\label{eqn:A-optimal-const-rel-e}
\end{eqnarray}
where $\lambda = N \left(\sum_{i=1}^m \left(\sqrt{m+1-i} - \sqrt{m-i}\right)
s_i\right)^{-1}$.  The minimum value is 
\begin{equation}
\min_{\{ n_e \}} \tr(C) = N^{-1} \left( \sum_{i=1}^m \left(\sqrt{m+1-i} - \sqrt{m-i}\right) s_i \right)^2.
\end{equation}
\label{cnj:A-optimal-const-rel-e}
\end{conjecture}

\begin{figure}
\includegraphics[width=2.5in]{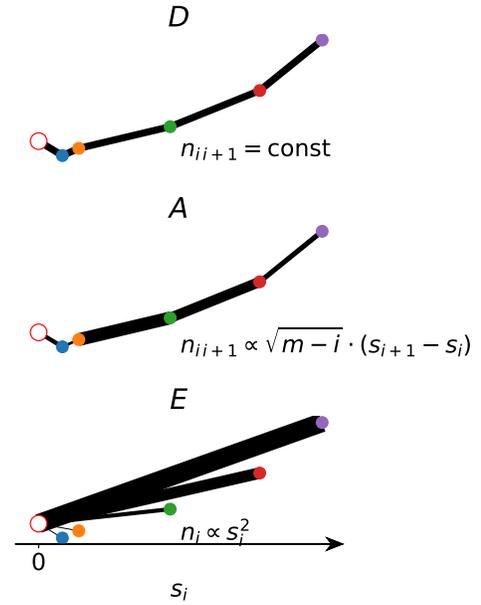}
\caption{The $D$-, $A$-, and $E$-optimals of difference networks where the
relative errors in the measurements are a constant.  The width of each edge
$e$ in the $\G_n$ is proportional to the allocation $n_e$. }
\label{fig:const-rel-e}
\end{figure}

In the $A$- and $D$-optimals, the subgraph of $\G_n$ consisting of only
the edges with weights $n_e \neq 0$ is the minimum spanning tree (MST)
of $\G_s$, connecting vertices $i$ and $i+1$, for $i=0,1,\dots,m-1$
(Fig.~\ref{fig:const-rel-e}).  The above conjectures have been
corroborated by comparing the constructive solutions with the results
of numerical minimizations for many sets of randomly generated $\{ s_e
\}$, but their rigorous proofs have so far defied me.

\section{Iterative optimization of the difference network}
\label{sec:iteration}

Better estimates of $\{ s_{e} \}$ are obtained as measurements
proceed, which can be used to derive improved allocations $\{ n_e
\}$.  Such a measurement-allocation cycle can be iterated so that the
resulting allocations in the difference network approach the true
optimal.  

Let $\{ n_e \}$ be the resources already spent to perform the
measurements, and $\{ s_e \}$ be the current estimates of the
fluctuations.  A total of $\Delta N$ new resources are to be allocated
to the measurements in the next iteration. It is straigthforward to
optimize the objective functions ({\it e.g.} $\tr(\C)$ in the
$A$-optimal) with respect to $\{ \Delta n_e \}$, where $\Delta n_e$ is
the additional resource to be allocated to measurement $e$, under
the constraints
\begin{equation}
\Delta n_e \geq 0 \text{ for all }e\text{, and }\sum_e \Delta n_e = \Delta N.
\end{equation}

For $A$-optimal, this minimization can again be cast as an SDP,
replacing $\F(\{ n_e \})$ in Eq.~\ref{eqn:dual-A-optimal} with $\F(\{
n_e + \Delta n_e \})$.

\section{Sparse difference network}
\label{sec:sparse-diffnet}

The following outlines a heuristic approach to designing
a sparse, near-optimal difference network, which is $k$-edge-connected
and entails $M$ measurements, {\it i.e.} $|\edges(\G_n)| = M$,
where $|\edges(\G_n)|$ is the number of edges with $n_e > 0$ in
the difference network $\G_n$.

\begin{enumerate}

\item Find the $k$-edge-connected spanning subgraph $\G_k$ of
  $\G_s$, such that the set of selected edges $E =
  \edges(\G_k)$ minimizes $\sum_{e\in E} s_e$.  For $k\leq
  2$, this problem can be solved in polynomial time; for $k>2$, an
  approximate solution can be found in polynomial
  time~\cite{Czumaj1999}.

\item Augment $\G_k$ with $M - |\edges(\G_k)|$ edges in
  $\edges(\G_s) \setminus \edges(\G_k)$ with the smallest $s_e$
  values, where $\edges(\G_s) \setminus \edges(\G_k)$ denotes
  the set of edges that are in $\G_s$ but not in $\G_k$.  The resulting graph
  $\G_{k,M}$ has $M$ edges and is $k$-edge-connected.

\item Setting $n_e=0$ for $e\not\in \edges(\G_{k,M})$, optimize
  the objective function with respect to $\{n_e|e\in
  \edges(\G_{k,M})\}$.

\end{enumerate}

\providecommand{\latin}[1]{#1}
\providecommand*\mcitethebibliography{\thebibliography}
\csname @ifundefined\endcsname{endmcitethebibliography}
  {\let\endmcitethebibliography\endthebibliography}{}

\end{document}